\documentclass[10pt, twocolumn, letterpaper, journal]{IEEEtran}
\usepackage{amssymb,amsmath,epsfig,tabularx,delarray,enumerate,subfigure,graphicx,array,cite,color}
\usepackage{tikz,mathpazo}
\usetikzlibrary{shapes.geometric,arrows}
\usepackage{chemarr,float,array,booktabs,multirow,threeparttable}
\usepackage[english]{babel}

\begin{document}
\title{FMO Interaction Energy between 17$\beta$-Estradiol, 17$\alpha$-Estradiol and Human Estrogen Receptor $\alpha$}
\author{Ricardo Ugarte$^\ast$
\thanks{$^\ast$Instituto de Ciencias Quimicas, Facultad de Ciencias, Universidad Austral de Chile. Independencia 641, Valdivia, Chile (e-mail: rugarte@uach.cl).}}
\markboth{december 2020}{}
\maketitle

\begin{abstract}

The estrogen receptor is a nuclear hormone receptor activated by the natural steroid hormone 17$\beta$-estradiol (E2). Fragment molecular orbital (FMO) calculations were performed which allowed us to obtain the interaction energy ($E_{int}$) between E2, 17$\alpha$-estradiol (17$\alpha$-E2) and the human estrogen receptor $\alpha$ ligand-binding domain. In aqueous media the MP2/6-31G(d) $E_{int}$ was of -88.52 kcal/mol for E2 and -78.73 kcal/mol for 17$\alpha$-E2. Attractive dispersion interactions were observed between ligands and all surrounding hydrophobic residues. Water molecules were found at the binding site and strong attractive electrostatic interactions were observed between the ligands and the Glu 353 and His 524 residues. The essential dynamics revealed that E2 adapts to the binding site and its motion, in a sense, synchronizes with the whole receptor; while 17$\alpha$-E2, with its motion of greater amplitude compared to E2, disturbs the binding site. Perhaps this feature of the normal substrate is a necessary condition for biological function. Another important requirement relates to the number of water molecules at the binding site. Therefore, negative values in $E_{int}$ is a necessary but not sufficient condition since, it is also necessary to consider the conformers population that fulfill all the requirements that ensure a biological response.

\end{abstract}


\section{Introduction}

The estrogen receptor (ER) is a ligand-inducible transcription factor and a member of the nuclear receptor superfamily that responds to endogenous (endobiotics) and exogenous (xenobiotics) chemicals by regulating gene expression \cite{1}. ER include two subtypes, ER$\alpha$ and ER$\beta$. These share structural patterns that are responsible for similar functional features. Both subtypes possess a modular organization that is characteristic of the nuclear receptors; five functional domains from the $ NH_{2}- $ to $ COOH $ terminus, designated A/B, C (DNA-binding domain or DBD), D, E (ligand-binding domain or LBD), and F \cite{2,3}. The full-length ER is not amenable to forming crystals make x-ray crystallography impossible.

There are three main mammalian endobiotics called estrogens: estrone (E1), 17$\beta$-estradiol (E2), and estriol (E3), all derived from cholesterol. All these estrogens act as ligands (L) for ER, but E2 (Figure 1) is the most abundant and physiologically active form as it exhibits extremely high binding affinity to ERs \cite{4}. 17$\alpha$-estradiol (17$\alpha$-E2) is a stereoisomer of E2 (C$_{17}$ epimer of E2), and according to somewhat contradictory data it would present an apparently lower estrogenic potency than E2 \cite{5,6,7,8}. The xenobiotics that interfere with the normal ER signaling include pharmaceuticals, industrial chemicals, pesticides, and phytoestrogens \cite{2}. Estradiol binding induces a major structural reorganization of the LBD that converts the inactive ER to the functionally active form by generating surfaces for enhanced stability of the ER dimer and of the interacting co-­regulatory proteins \cite{9}.

Numerous crystal structures have been determined for the LBDs of both subtypes, and these have given a detailed insight into the structure and alterations during the ligand binding. The structure of ER LBD reveals a conserved core of twelve $\alpha$-helices and a short two-stranded antiparallel $\beta$-sheet arranged into a three-layered sandwich fold; this arrangement generates a mostly hydrophobic cavity in the lower part of the domain which can accommodate the ligand \cite{1} (Figure 2).

A number of experimental and theoretical studies have been performed to investigate the L-ER interaction \cite{11,12,13,14,15,16,17,18,19,20,21,22,23,24,25,26,27,28,29,30,31,32,33,34}, and since 1997 about 100 crystal structures of ER LBD with several ligands have been solved and deposited in the Research Collaboratory for Structural Bioinformatics (RCSB) Protein Data Bank (PDB). On the basis of the above information, the mode of binding between ERs and their ligands has been determined. The specific recognition between ER and its ligand mainly depends on hydrogen bonds and hydrophobic contacts \cite{35,36}.

Most of the theoretical studies which use the structures deposited in RCSB PDB are carried out using molecular dynamics (MD) simulations. These simulations are based on empirical force fields that may not be accurate enough to predict L-ER interaction energies. Accuracy requirements could be provided by ab initio quantum mechanical calculations, but these can be very computationally expensive and time consuming. The hybrid QM/MM (quantum mechanics/molecular mechanics) is a method that combines the precision of quantum mechanics and the speed of empirical force fields. In this approach, part of the system that includes the chemically relevant region is treated quantum mechanically (QM) while the remainder, often referred to as the environment, is treated at the classical level using empirical or molecular mechanics (MM) force fields. This multiscale approach reduces the computational cost significantly as compared to a QM treatment of the entire system and makes simulations possible \cite{37,38}.

\begin{figure} [!htp]
\begin{center}
\includegraphics[width=0.7\columnwidth]{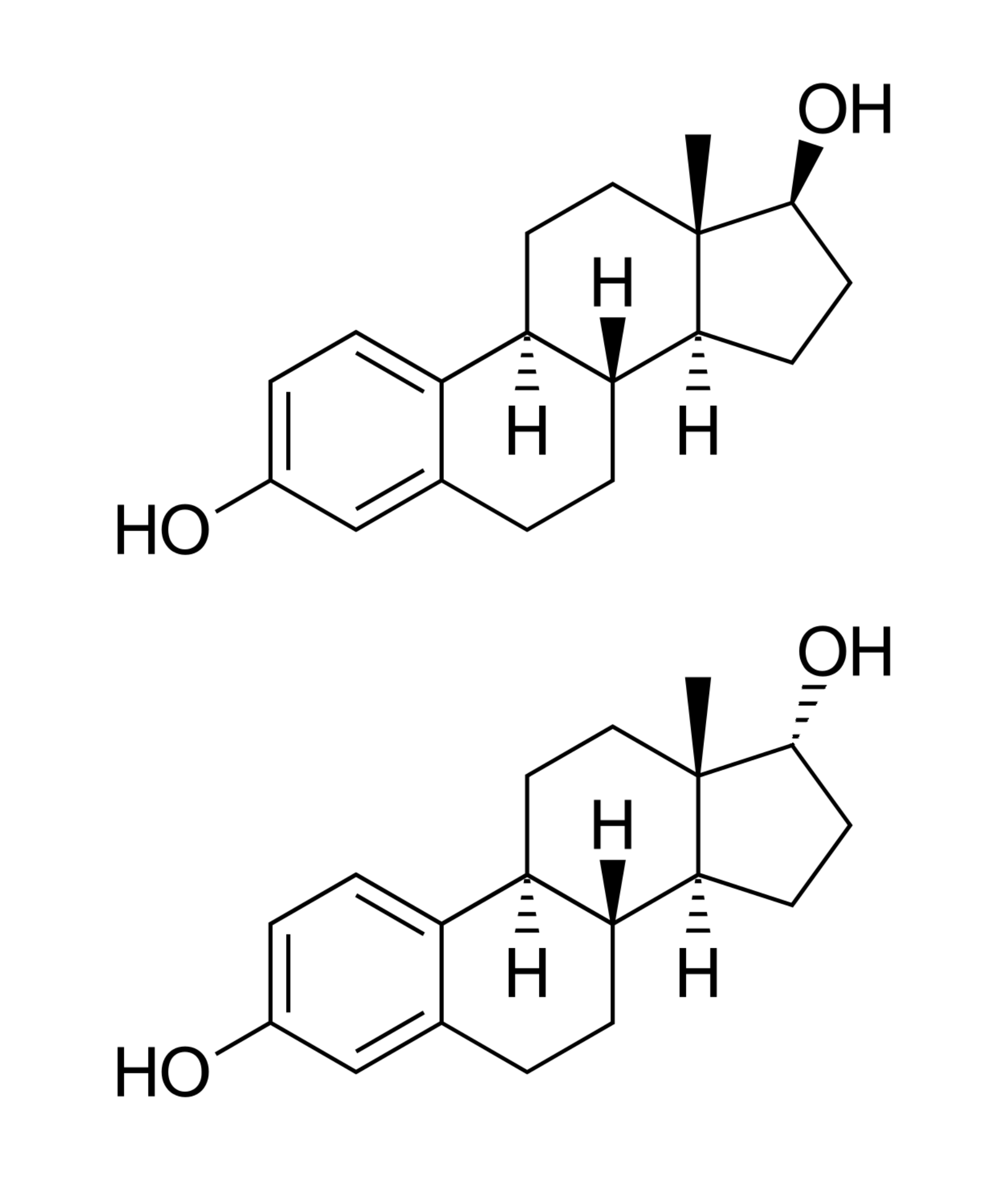}
\end{center}
    \caption{Chemical structure of 17$\beta$-Estradiol (top) and 17$\alpha$-Estradiol (bottom).}
    \label{Figure1}
\end{figure}

\begin{figure} [!htp]
\begin{center}
\includegraphics[width=1.0\columnwidth]{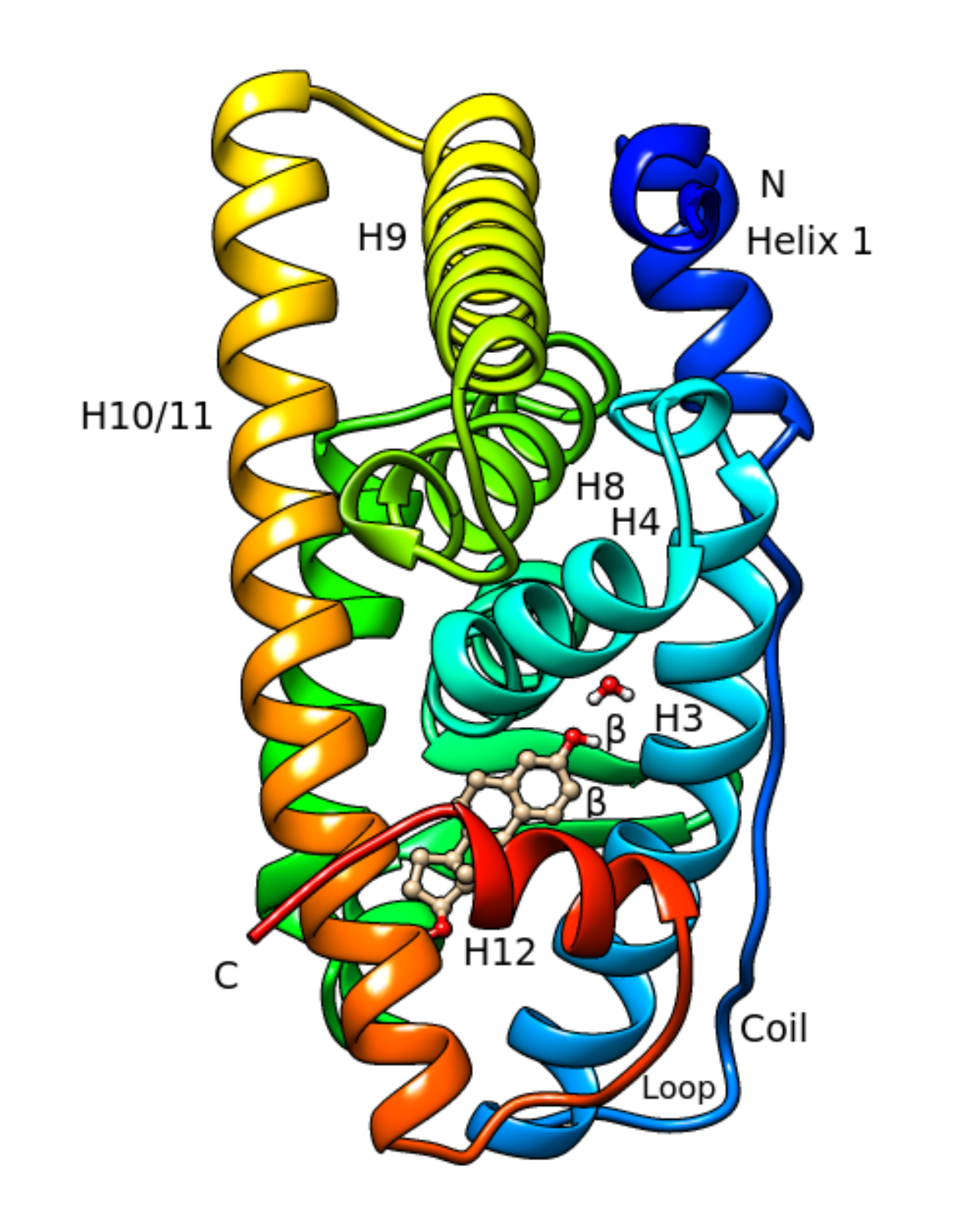}
\end{center}
    \caption{Model of HER$\alpha$ LBD (ribbon). E2 and water molecule (ball and stick) at the binding site \cite{10}. The model based on the RSCB PDB crystal structure (PDB code 1G50) includes 247 amino acid residues.}
    \label{Figure2}
\end{figure}

An efficient alternative to either the full ab initio QM, MM or QM/MM, lies in the fragment-based methods, which form an actively developed field of research \cite{39}. Fragment Molecular Orbital (FMO) method \cite{40,41} is one such method that has been used for efficient and accurate QM calculations in very large molecular systems \cite{21,42}. FMO involve fragmentation of the chemical system, and from ab initio or density functional quantum-mechanical calculations of each fragments (monomers) and their dimers (and trimers if greater accuracy is required) one can construct the total properties. The distinctive feature of FMO is the inclusion of the electrostatic field from the whole system in each individual fragment calculation, and in using the systematic many-body expansion. The FMO method is suited to various analyses, as it provides information on fragments and their interactions that are naturally built into the method.

In the present article, we report a study on an calculation of the interaction energy between both
estradiol isomers and LBD of human estrogen receptor alpha in aqueous medium. The aim is to further explore the mechanism of interaction underlying to the L-ER binding. In particular, verify if the calculation method is capable of discriminating between two very similar compounds, which bind to the same binding site. Briefly, the main steps of the calculation are as follows: (i) Search for representative structures of the conformational space around the crystallographic state of L-ER$\alpha$ LBD by means of MD simulations and cluster analysis. (ii) Geometry optimization of the representative structures using QM/MM approach. (iii) Single point FMO calculation on the above optimized structures in order to obtain the inter-fragment interaction energies.

\section{Methods}

\subsection{Model Building}

Crystal structure of the HER$\alpha$ LBD in complex with E2 (PDB code 1G50) at 2.9 {\AA} resolution was retrieved from the RSCB PDB \cite{43}. The model was built from chain A of the homodimer. Missing hydrogen atoms were added with the LEaP module of AmberTools 15 package \cite{44}. AMBER FF14SB force field was selected for the proteins and general AMBER force field (GAFF) parameters \cite{45} were employed for E2. In order to parameterize E2, electrostatic potential was calculated by Gaussian 09 program at the HF/6-31G(d) level of theory \cite{46}. Partial charges were fitted by RESP method of the Antechamber module of AmberTools 15 \cite{47}. Asp 313, Asp 321 and Glu 323 were modeled as neutral species, Arg, Lys and remainder of Asp, Glu residues as charged species, all tyrosines as neutral. Three of the 13 residues of histidine (His 476, His 501, His 513), were protonated in order to preserve the electroneutrality of the system. The N- and C-terminus residues were protonated and deprotonated, respectively. The model was solvated with TIP3P water in a pre-equilibrated box measuring 89 x 78 x 78 {\AA}$^{3}$. The E2-ER-W system contains 247 amino acid residues (4009 atoms), E2 and 13394 water molecules (W). The total number of atoms in the system is 44235.

Finally, the same previous protocol was used to build 17$\alpha$-E2-ER-W and ER-W. The molecular structure of 17$\alpha$-E2 in the binding site was derived from E2 by modifying the orientation of the corresponding OH group, while the ER-W structure was generated by eliminating the ligand.

\subsection{Molecular Dynamics Simulations} 

In order to remove the steric strain introduced when adding hydrogen atoms, holo (E2-ER-W, 17$\alpha$-E2-ER-W) and apo (ER-W) models were subjected to three successive steps of minimization using the SANDER module of the AmberTools 15. First, 1000 steps of steepest descent followed of 1500 steps of conjugate gradient, allowing only H atoms and water to move while holding the rest of the system fixed. Next, the same minimization algorithm is used as the previous one, but maintaining the water fixed. Finally, the whole system was minimized without any restraints for 2000 steps of steepest descent followed by 1000 steps of conjugate gradient.

All simulations were carried out with the SANDER module of the AmberTools 15 with periodic boundary conditions, using Particle Mesh Ewald method \cite{48} to treat long-range electrostatics interactions with a non-bonded cutoff of 10 {\AA}. All bonds involving hydrogen atoms were restrained using the SHAKE \cite{49} algorithm. Temperature regulation was done using a Langevin thermostat with collision frequency of 1 ps$^{-1}$. The Berendsen barostat was used for constant pressure simulation at 1 atm, with a relaxation time of 1 ps. The time step was 1 or 2 fs. The hydrogen mass distribution (HMR) method was used for accelerating MD simulations \cite{50}. The final energy-minimized models was then submitted to the following protocol:
\\

\tikzstyle{startstop} = [rectangle, rounded corners, minimum width=3cm, minimum height=1cm,text centered, draw=black, fill=red!30]
\tikzstyle{arrow} = [->,>=stealth]

Scheme 1:
\\
\begin{center}
\begin{tikzpicture}[node distance=2cm]
\node (start) [startstop] {NVT: 0 $\rightarrow$ 310 K \ $ \Delta t = 1 $ fs \ 100 ps};
\node (in1) [startstop, below of=start] {NPT: 310 K \ $ \Delta t = 2 $ fs \ 500 ps};
\node (pro1) [startstop, below of=in1] {NVT: 310 $\rightarrow$ 5 K \ $ \Delta t = 1 $ fs \ 100 ps};
\draw [arrow](start) -- (in1);
\draw [arrow](in1) -- (pro1);
\end{tikzpicture}
\end{center}
$  $

From the restart file of the last simulation in Scheme 1, we conducted an extensive set of molecular dynamics simulations to explore the conformational space in the vicinity of the crystallographic structure. To circumvent the limited conformational sampling ability of MD simulations at 310 K, we used multiple-trajectory short-time simulations \cite{51}. By combining the sampling ability of the multiple trajectories, we expect to sample more conformational space than single trajectory of the same length. The aforementioned restart file was used as seed for 30 short-time simulations that obey the protocol established in Scheme 2:
\\

Scheme 2:
\\
\begin{center}
\begin{tikzpicture}[node distance=2cm]
\node (start) [startstop] {NVT: 5 $\rightarrow$ 150 K \ $ \Delta t = 2 $ fs \ 60 ps};
\node (in1) [startstop, below of=start] {NPT: 150 $\rightarrow$ 310 K \ $ \Delta t = 2 $ fs \ 140 ps};
\node (pro1) [startstop, below of=in1] {NVE: 310 K \ $ \Delta t = 2 $ fs \ 500 ps};
\draw [arrow](start) -- (in1);
\draw [arrow](in1) -- (pro1);
\end{tikzpicture}
\end{center}
$  $

The initial velocities (Scheme 2) were assigned randomly from a Maxwell-Boltzmann distribution at 5 K. The trajectories start with the same structure and differ only in the initial velocity assignment. At the end of the equilibration, from $\sim$ 80 ps NPT ensemble, the average temperature of the final 60 ps was 310 K, and the average density was 1.0 g/mL. All production runs of 0.5 ns were performed in an NVE ensemble at 310 K.

\subsection{Cluster Analysis}

\begin{table*}[!htbp]
 \begin{center}
  \begin{threeparttable}
\caption{Cluster Analysis of MD trajectories in the different models}
\begin{tabular}{cccccccccccccccc}
  \toprule[1pt]
  & \multicolumn{5}{c}{{E2-ER-W}}
  & \multicolumn{5}{c}{{17$\alpha$-E2-ER-W}}
  & \multicolumn{5}{c}{{ER-W}} \\
  \midrule[1pt]
    Cluster & 1 & 2 & 3 & 4 & 5 & \ \ \ 1 & 2 & 3 & 4 & 5 & \ \ \ 1 & 2 & 3 & 4 & 5 \\
  CP$^{(a)}$ & 3630 & 1749 & 1169 & 907 & 45 & \ \ \ 5156 & 185 & 543 & 77 & 1539 & \ \ \ 1800 & 3759 & 1652 & 191 & 98 \\
  CP \% & 48.4 & 23.3 & 15.6 & 12.1 & 0.6 & \ \ \ 68.8 & 2.5 & 7.2 & 1.0 & 20.5 & \ \ \ 24.0 & 50.1 & 22.0 & 2.6 & 1.3 \\
  RS$^{(b)}$ & \ $\beta$I & $\beta$II & $\beta$III & $\beta$IV & $\beta$V & \ \ \ $\alpha$I & $\alpha$II & $\alpha$III & $\alpha$IV & $\alpha$V & \ \ \ I & II & III & IV & V \\
  \bottomrule[1pt]
\end{tabular}
\begin{tablenotes}
      \footnotesize
      \item $^{(a)}$Cluster population. $^{(b)}$Representative structure of the respective cluster.
\end{tablenotes}
\label{Table I}
  \end{threeparttable}
    \end{center}
\end{table*}

Cluster analysis methods have been developed for analyzing simulation trajectories of biomacromolecules and used for the analysis of their conformational behavior in solution \cite{52,53}. These methods group together similar conformers from molecular simulations. A clustering approach for each model, based on the C$\alpha$-RMSD (root mean square deviation) was applied to the snapshots of the MD simulations \cite{54}. We selected the alpha carbon atoms because they describe the backbone conformations. The C$\alpha$-RMSD was calculated after rigid body alignment of C$\alpha$ atoms of each frame of the trajectory with respect to C$\alpha$ atoms of the average structure of the protein. Prior to clustering, the individual trajectories from the 30 short-time simulations were combined into a single file (full trajectory), and the water molecules were removed to speed up the calculations. In our analysis, 7500 snapshots were grouped into five clusters. Each cluster is described by a centroid structure, which in itself is not physically significant as it is effectively a mathematical construct based on the members of a cluster. However, the actual structure closest to the centroid (rmsd) is significant and representative of each cluster. Thus, five representative structures (RS) of each cluster, and therefore of the conformers population, were obtained (Table I).

\subsection{QM/MM Optimization}

All L-ER-W models representative of the population were subjected to geometry optimization using Gaussian 09 at the MM/AMBER level of theory. Then, to facilitate the QM/MM calculations, the water molecules beyond 10 {\AA} of the protein surface were deleted using VMD program \cite{55}. As a consequence, new models (L-ER-w) with a water layer of 10 {\AA} around of receptor surface were generated. ONIOM, \cite{37} a hybrid QM/MM method implemented in Gaussian 09, was used for the geometry optimization of L-ER-w models. In the present study we used a two-layer ONIOM (B3LYP/6-31G(d):AMBER) scheme: L(B3LYP/6-31G(d)); ER-w(AMBER).

\subsection{FMO Calculations}

Fragment-based approaches refer to the chemical idea of parts of the system retaining their identity to a large extent (e.g., functional groups and residues). FMO method not only reduces the computational cost, but it also provides a wealth of information on the properties of fragments and their interactions. The calculation of the ligand-receptor interaction energy is based on obtaining and sum the interaction energies between all pairs of fragments ligand-amino acid residue and if applicable, ligand-water trapped on binding site.

\begin{table}
 \begin{center}
 \begin{threeparttable}
\caption{Number of FMO fragments in the L-ER-w models$^{(a)}$}
 \begin{tabular}{ccc}
 \toprule[1pt] 
    Fragments & E2-ER-w & 17$\alpha$-E2-ER-w \\
     \midrule[1pt]
 17$\beta$-estradiol & 1 & \\
 17$\alpha$-estradiol & & 1 \\
   Amino acid residue & 246 & 246 \\
          Water & 4179 & 4179 \\
 \midrule[1pt]
          Total & 4426 & 4426 \\ 
 \bottomrule[1pt]
 \end{tabular}
     \begin{tablenotes}
      \footnotesize
      \item $^{(a)}$Models with a water layer of 10 {\AA} around of receptor surface.
    \end{tablenotes}
\label{Table II}
  \end{threeparttable}
  \end{center}
\end{table}

The AFO (adaptive frozen orbitals) scheme was used throughout for fragmentation across peptide bonds, with the default settings for bond definitions. The fragmentation of the model was as follows: The first two amino acid residues and each remaining amino acid residue of apo-ER, L, and the water molecule were treated as a single fragment. Table II shows the number of fragments in the different models. In order to reduce the computational cost, multilayer two-body FMO calculation were performed. The molecular system was divided into two layers treated at different levels of theory: FMO2-RHF/STO-3G:MP2/6-31G(d). The layer 1 (aqueous environment) described by RHF/STO-3G and layer 2 (L-ER) described by MP2/6-31G(d) \cite{56}. The water molecules of the binding site were included in the layer 2. 

To contrast the interaction energies, models in vacuo were built from the original models in which all water molecules were removed, except those found on the binding site; thus, these water molecules are incorporated into the receptor. The E2-ER and 17$\alpha$-E2-ER were subjected to FMO2-MP2/6-31G(d) calculations. 

Finally, pair interaction energies were computed in the following models: E2-ER-w, 17$\alpha$-E2-ER-w, E2-ER and 17$\alpha$-E2-ER.

\section{Results and Discussion}

\subsection{Analysis of Molecular Dynamics Trajectories}

The RMSD with respect to average structure or RMSF were calculated ({\AA}) for all three models: apo-ER: 0.76 $ \pm $ 0.12; E2-ER: 0.75 $ \pm $ 0.10; 17$\alpha$-E2-ER: 0.80 $ \pm $ 0.12, which indicates that the simulations represent fluctuations around a stable average. The RMSD value of the holo-ER average structures with respect to the apo-ER average structure taken as reference are 0.98 {\AA} and 0.72 {\AA} for E2-ER and 17$\alpha$-E2-ER, respectively (Figure 3). Apparently 17$\beta$-estradiol affects the apo-ER backbone more than 17$\alpha$-estradiol. Detailed observation of the structures shows a slight increase of random coil in holo-ER with respect to apo-ER (the increase is higher in E2-ER). The above, is in detriment of the content of alpha helix. The following motifs are not influenced by the ligands binding: H1/H2, Coil, H3, H6 and H10/H11. The loop H3-H4 is only affected by 17$\alpha$-E2 binding; the loop H8-H9 is significantly affected by E2, and only slightly affected by 17$\alpha$-E2. The N- and C-terminus are also slightly modified by ligands binding, although to a larger extent by E2. The $\beta$-hairpins is modified in the same manner by the ligands binding. With regard to the ligands, both are placed in the same position on the binding site (Figure 4).

\begin{figure*} [!htp]
\begin{center}
\includegraphics[width=1.0\linewidth]{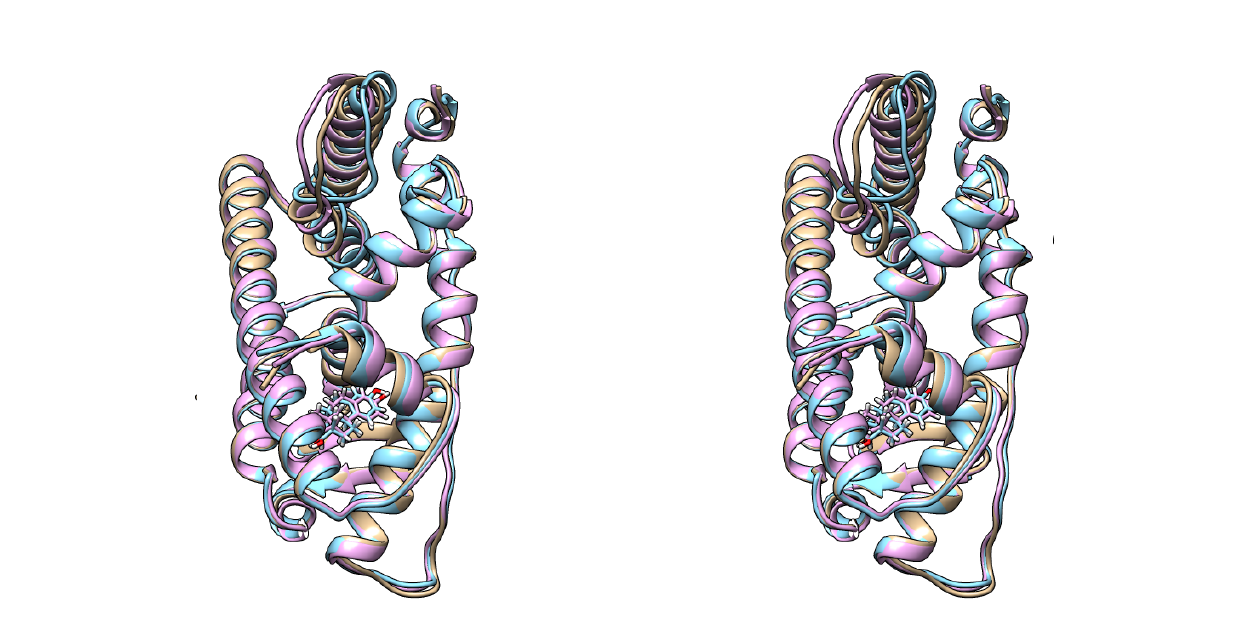}
\end{center}
    \caption{Stereo view of the average structures over the trajectories: apo-ER (tan); E2-ER (sky blue) and 17$\alpha$-E2-ER (pale plum). Estrogen Receptor: ribbon; Ligands: ball and stick. Structure superposition based on the C$\alpha$. Taking as a reference the apo-ER structure, the root mean square distance (RMSD) value for E2-ER and 17$\alpha$-E2-ER is 0.98 {\AA} and 0.72 {\AA}, respectively.}
    \label{Figure3}    
\end{figure*}

\begin{figure} [!htp]
\begin{center}
\includegraphics[width=1.0\columnwidth]{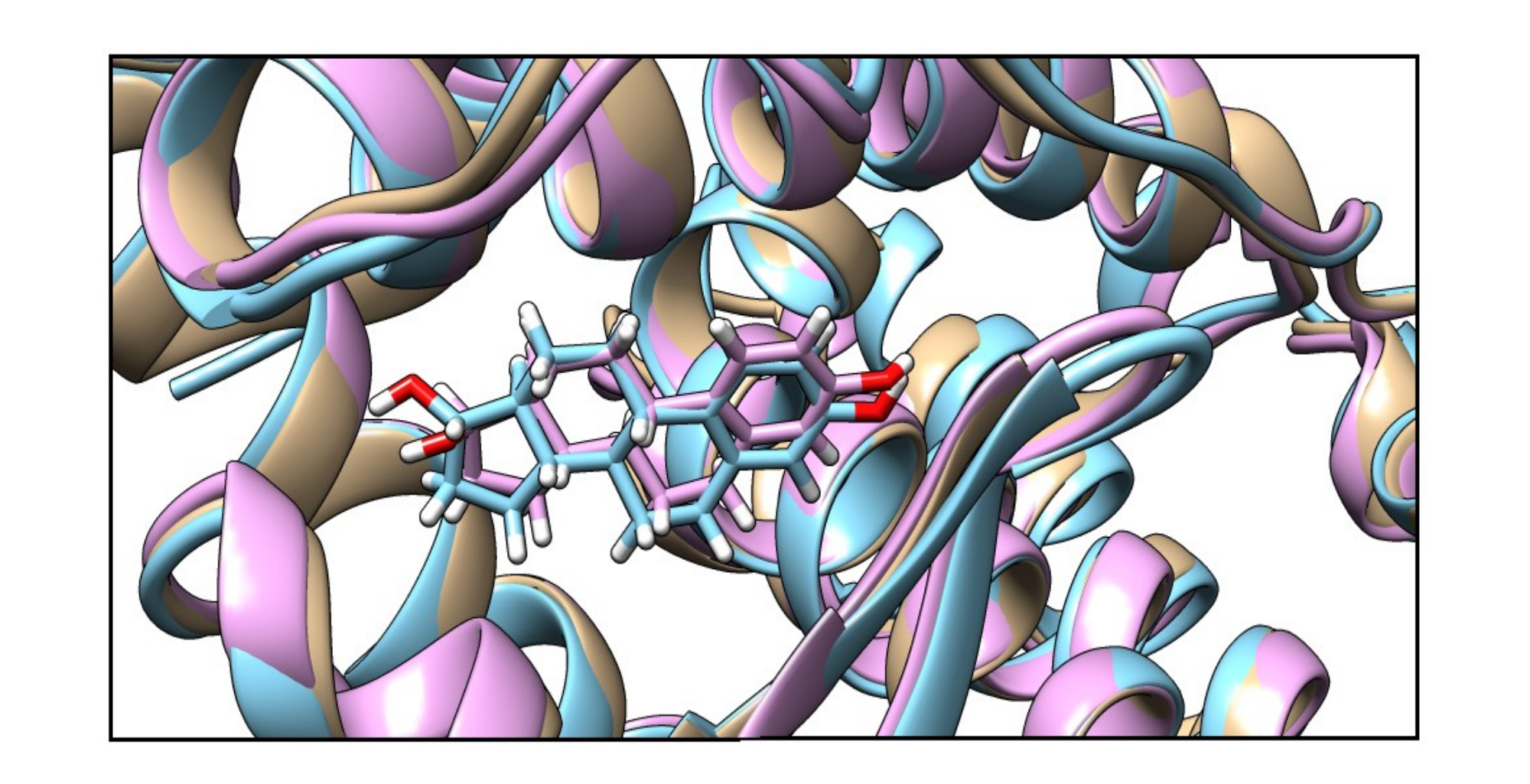}
\end{center}
    \caption{17$\beta$-Estradiol and 17$\alpha$-Estradiol on the binding site. Detail from image taken from the Figure 3.}
    \label{Figure4}
\end{figure}

In order to compare the dynamic behavior of the three models, the essential dynamics method was used \cite{57,58}. For this purpose, the following procedure was carried out: (i) Combination of the full trajectories (water stripped) into a single trajectory (22500 frames) and removal of ligand molecules. (ii) Principal component analysis (PCA) \cite{59} on the combined MD trajectory to calculate a set of eigenvectors, and then projected the backbone C$\alpha$ fluctuations onto these eigenvectors or principal components. (iii) Clustering of the combined MD trajectory and PCA data sorted by cluster \cite{54}.

\begin{figure*} [!htp]
\begin{center}
\includegraphics[width=0.8\linewidth]{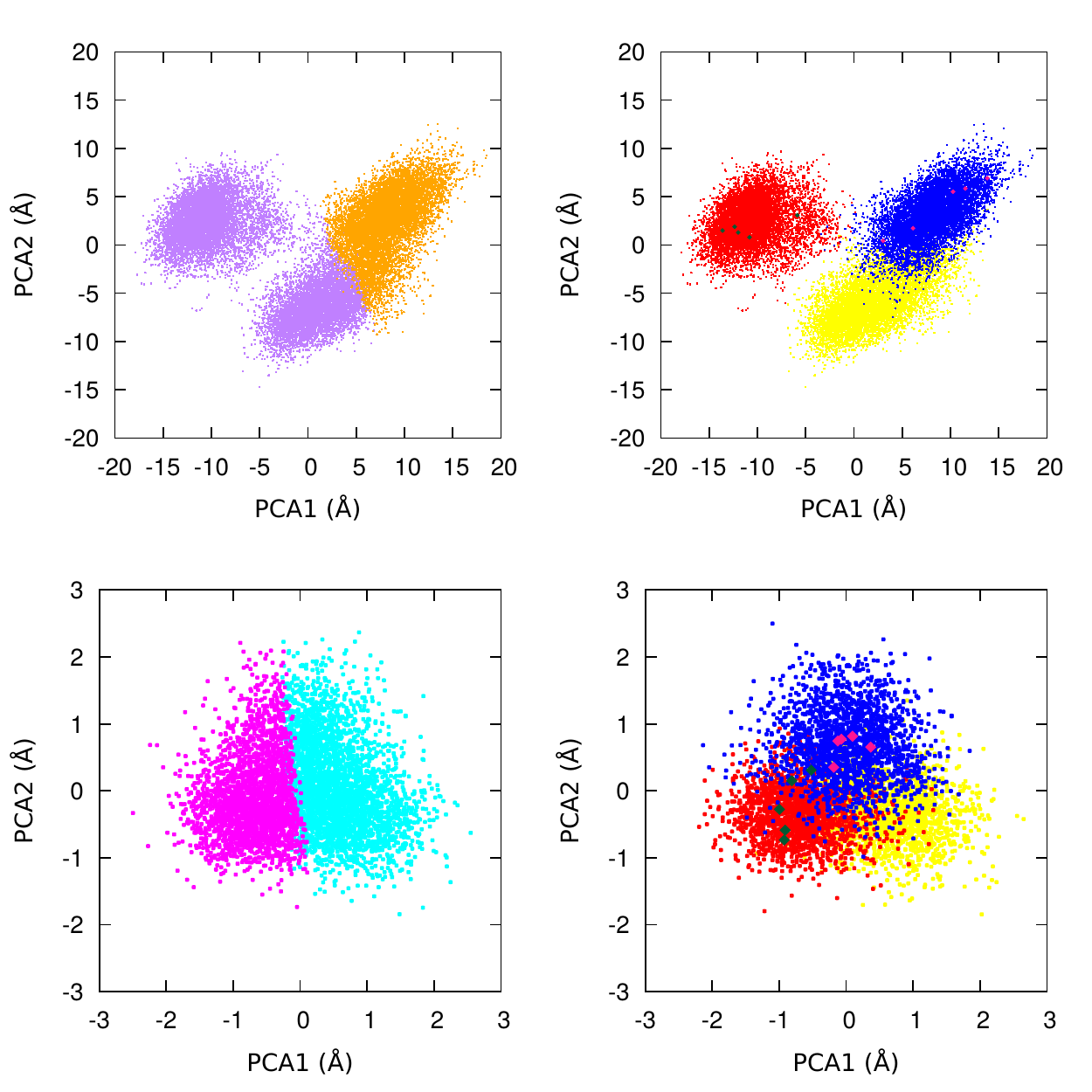}
\end{center}
    \caption{Clustering results of MD simulations projected on 2D plane formed by first two PCs. Clustering was performed on the combined MD trajectory data and two clusters were requested in the computation. Cluster separation for the backbone dynamics of the entire estrogen receptor (top left) and its binding site (bottom left): cluster 1r$^{a}$ (purple); cluster 2r (orange); cluster 1bs$^{b}$ (magenta); cluster 2bs (cyan). Distribution of motion into the clusters (top and bottom right): ER (yellow); E2-ER (red); 17$\alpha$-E2-ER (blue); E2-ER (diamond dark green) and 17$\alpha$-E2-ER (diamond dark pink) representative structures. \ $^{a}$r: receptor; $^{b}$bs: binding site.}
\label{Figure5}
\end{figure*}

\begin{figure} [!htp]
\begin{center}
\includegraphics[width=0.6\columnwidth]{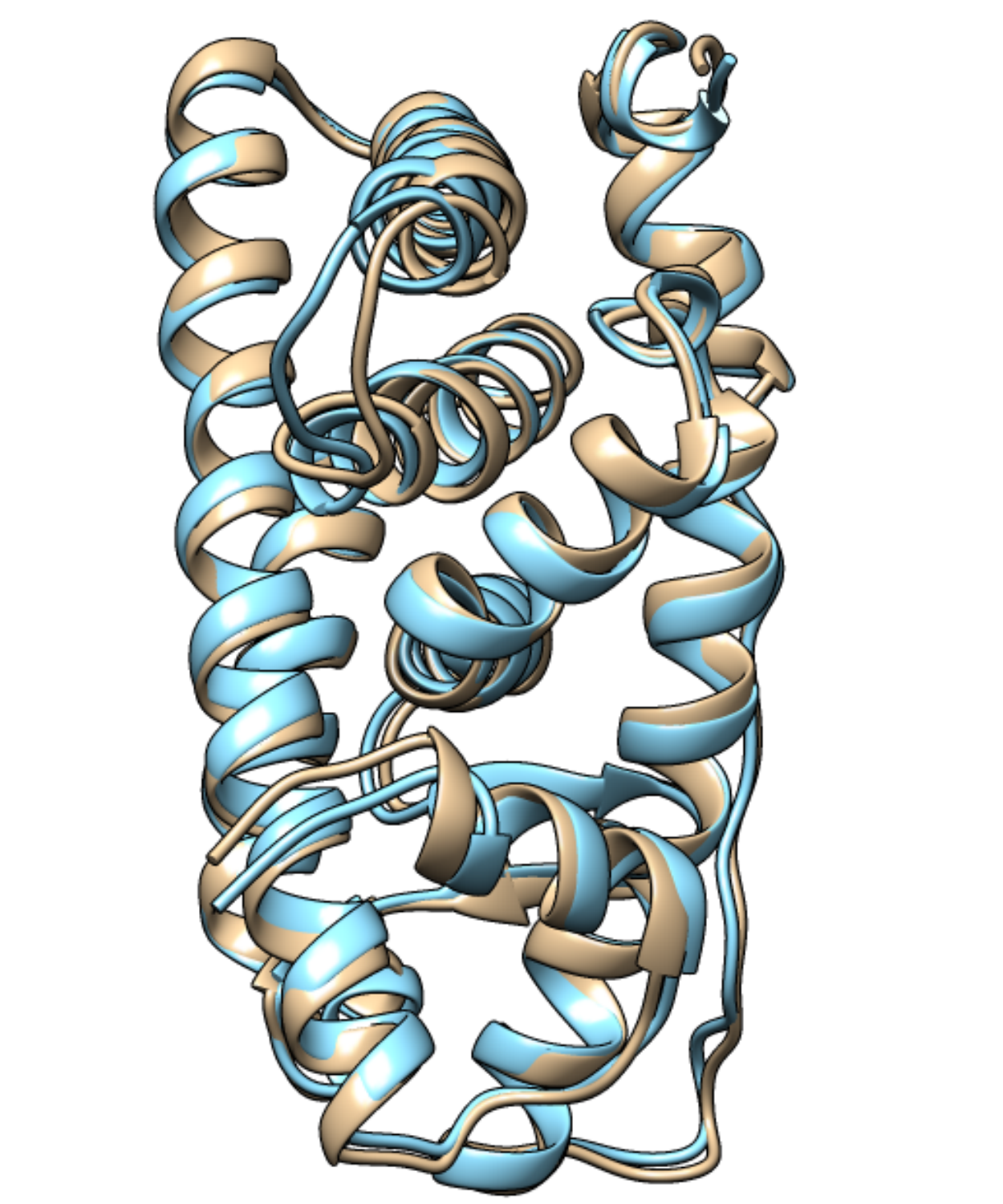}
\end{center}
    \caption{Superposition of representative structures from cluster 1r (tan) and 2r (sky blue). The root mean square distance (RMSD) value for this superposition is 1.045 {\AA}.}
    \label{Figure6}
\end{figure}

\begin{figure*} [!htp]
\begin{center}
\includegraphics[width=0.95\linewidth]{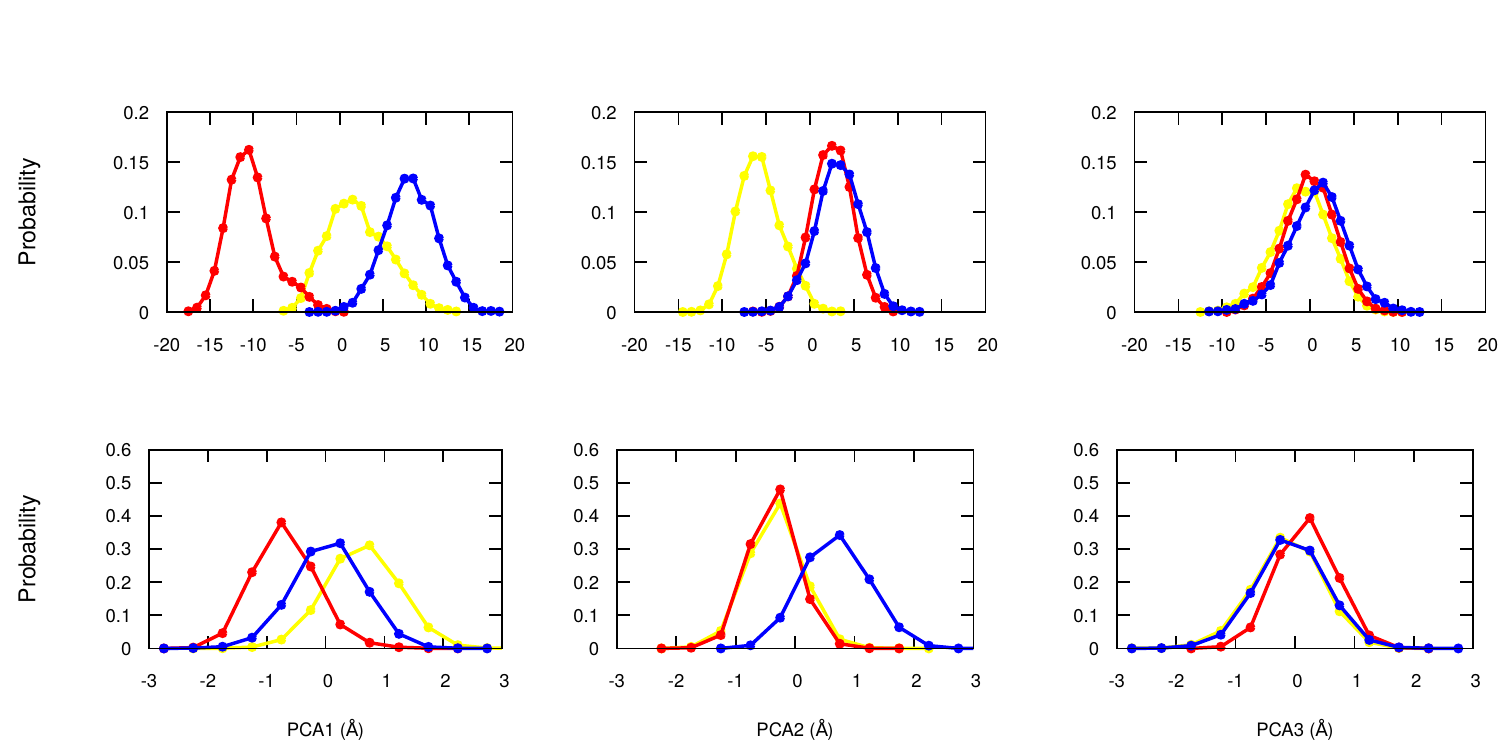}
\end{center}
    \caption{Probability distributions for the backbone C$\alpha$ fluctuations in combined MD trajectory along several principal components. The entire estrogen receptor (top) and its binding site (bottom): apo-ER (yellow); E2-ER (red); 17$\alpha$-E2-ER (blue).} 
    \label{Figure7}
\end{figure*}

While PCA reveals the essential dynamics or main motions (large-scale motion) contained in an MD trajectory it does not partition the frames into distinct conformational categories. This can be achieved by clustering the PC data. A previous study confirms the choice of two clusters. If three clusters are requested, the third represents a population percentage of approximately 0.5\%. The trajectory of the atomic displacements or fluctuations described by principal components (PCs) 1 and 2 are shown in Figure 5. Clustering for the C$\alpha$ fluctuations of the entire estrogen receptor is displayed in figure 5 (top left). This trajectory contains two conformational categories, practically separated along the first PC. Distribution of motion of apo-ER and holo-ER into the clusters are shown (top right). Interestingly, the apo-ER form displays dynamics characteristics of both clusters. The representative structure of cluster 1r belongs to the E2-ER population, while that of cluster 2r belongs to the 17$\alpha$-E2-ER population (Figure 6). The motions along these PCs are qualitatively similars since the point cloud represented in the plane has analogous sizes and shapes for apo-ER and holo-ER forms. This seems to indicate that there is no significant change in global dynamics of estrogen receptor upon uptake of ligand. However, the simulations visit different areas of the essential 2D space, indicating that there may be a small difference in the conformations associated with each model. Remarkably, a subset of trajectories of apo-ER and 17$\alpha$-E2-ER overlap; apparently, the overall differences between these forms are not as pronounced.

\begin{table*}[!htbp]
 \begin{center}
  \begin{threeparttable}
\caption{FMO Interaction Energy}
\begin{tabular}{cccccccccccc}
  \toprule[1pt]
       \multicolumn{11}{c}{In Aqueous FMO2-RHF/STO-3G:MP2/6-31G(d)} \\
  \midrule[1pt]
  & \multicolumn{5}{c}{{E2-ER-w}}
  & \multicolumn{5}{c}{{17$\alpha$-E2-ER-w}} \\
    \midrule[1pt]
  \ \ \ RS & \multicolumn{1}{c}{$\beta$I} & \multicolumn{1}{c}{$\beta$II} & \multicolumn{1}{c}{$\beta$III} & \multicolumn{1}{c}{$\beta$IV} & \multicolumn{1}{c}{$\beta$V} & \multicolumn{1}{c}{$\alpha$I} & \multicolumn{1}{c}{$\alpha$II} & \multicolumn{1}{c}{$\alpha$III} & \multicolumn{1}{c}{$\alpha$IV} & \multicolumn{1}{c}{$\alpha$V} \\
$E_{L-Aa}$ & \multicolumn{1}{c}{-88.47} & \multicolumn{1}{c}{-82.82} & \multicolumn{1}{c}{-82.91} & \multicolumn{1}{c}{-82.43} & \multicolumn{1}{c}{-79.37} & \multicolumn{1}{c}{-77.54} & \multicolumn{1}{c}{-81.36} & \multicolumn{1}{c}{-75.18} & \multicolumn{1}{c}{-67.44} & \multicolumn{1}{c}{-81.73} \\  
$E_{L-w}$ & \multicolumn{1}{c}{-3.21} & \multicolumn{1}{c}{-0.70} & \multicolumn{1}{c}{-6.90} & \multicolumn{1}{c}{-1.50} & \multicolumn{1}{c}{-7.70} & \multicolumn{1}{c}{} & \multicolumn{1}{c}{-8.69} & \multicolumn{1}{c}{-6.62} & \multicolumn{1}{c}{-11.43} & \multicolumn{1}{c}{1.45} \\ 
$E_{int-RS}^{(a)}$ & \multicolumn{1}{c}{-91.68} & \multicolumn{1}{c}{-83.52} & \multicolumn{1}{c}{-89.81} & \multicolumn{1}{c}{-83.93} & \multicolumn{1}{c}{-87.07} & \multicolumn{1}{c}{-77.54} & \multicolumn{1}{c}{-90.05} & \multicolumn{1}{c}{-81.80} & \multicolumn{1}{c}{-78.87} & \multicolumn{1}{c}{-80.28} \\ 
$E_{int}^{(b)}$ & \multicolumn{5}{c}{-88.52} & \multicolumn{5}{c}{-78.73} \\ 
\midrule[1pt]
\\
\\
\midrule[1pt] 
       \multicolumn{11}{c}{In Vacuo FMO2-MP2/6-31G(d)} \\
  \midrule[1pt]
  & \multicolumn{5}{c}{{E2-ER}}
  & \multicolumn{5}{c}{{17$\alpha$-E2-ER}} \\
    \midrule[1pt]
  \ \ \ RS & \multicolumn{1}{c}{$\beta$I} & \multicolumn{1}{c}{$\beta$II} & \multicolumn{1}{c}{$\beta$III} & \multicolumn{1}{c}{$\beta$IV} & \multicolumn{1}{c}{$\beta$V} & \multicolumn{1}{c}{$\alpha$I} & \multicolumn{1}{c}{$\alpha$II} & \multicolumn{1}{c}{$\alpha$III} & \multicolumn{1}{c}{$\alpha$IV} & \multicolumn{1}{c}{$\alpha$V} \\
$E_{L-Aa}$ & \multicolumn{1}{c}{-91.80} & \multicolumn{1}{c}{-86.34} & \multicolumn{1}{c}{-87.22} & \multicolumn{1}{c}{-85.47} & \multicolumn{1}{c}{-80.78} & \multicolumn{1}{c}{-80.93} & \multicolumn{1}{c}{-82.84} & \multicolumn{1}{c}{-79.94} & \multicolumn{1}{c}{-69.16} & \multicolumn{1}{c}{-84.40} \\  
$E_{L-w}$ & \multicolumn{1}{c}{-2.829} & \multicolumn{1}{c}{-0.755} & \multicolumn{1}{c}{-6.256} & \multicolumn{1}{c}{-1.140} & \multicolumn{1}{c}{-7.925} & \multicolumn{1}{c}{} & \multicolumn{1}{c}{-8.784} & \multicolumn{1}{c}{-6.110} & \multicolumn{1}{c}{-11.78} & \multicolumn{1}{c}{1.37} \\ 
$E_{int-RS}^{(a)}$ & \multicolumn{1}{c}{-94.63} & \multicolumn{1}{c}{-87.10} & \multicolumn{1}{c}{-93.48} & \multicolumn{1}{c}{-86.61} & \multicolumn{1}{c}{-88.70} & \multicolumn{1}{c}{-80.93} & \multicolumn{1}{c}{-91.62} & \multicolumn{1}{c}{-86.05} & \multicolumn{1}{c}{-80.94} & \multicolumn{1}{c}{-83.03} \\ 
$E_{int}^{(b)}$ & \multicolumn{5}{c}{-91.69} & \multicolumn{5}{c}{-82.00} \\  
  \bottomrule[1pt]   
\end{tabular}
\begin{tablenotes}
      \footnotesize
      \item $L$: Ligand; $Aa$: Amino acids; $w$: waters. $^{(a)}$ $E_{int-RS} = (E_{L-Aa} + E_{L-w})$. $^{(b)}$Represents the ligand-receptor interaction energy and correspond to the average weighted energy based on the cluster population. All energies at MP2/6-31G(d) level of theory in kcal/mol.
\end{tablenotes}
\label{Table III}
  \end{threeparttable}
      \end{center}
\end{table*}

\begin{table*}[!htbp]
 \begin{center}
  \begin{threeparttable}
\caption{FMO Interaction Energy}
\begin{tabular}{cccccccccccc}
  \toprule[1pt]
       \multicolumn{11}{c}{In Aqueous FMO2-RHF/STO-3G:MP2/6-31G(d)} \\
  \midrule[1pt]
  & \multicolumn{5}{c}{{E2-ER-w}}
  & \multicolumn{5}{c}{{17$\alpha$-E2-ER-w}} \\
    \midrule[1pt]
  \ \ \ RS & \multicolumn{1}{c}{$\beta$I} & \multicolumn{1}{c}{$\beta$II} & \multicolumn{1}{c}{$\beta$III} & \multicolumn{1}{c}{$\beta$IV} & \multicolumn{1}{c}{$\beta$V} & \multicolumn{1}{c}{$\alpha$I} & \multicolumn{1}{c}{$\alpha$II} & \multicolumn{1}{c}{$\alpha$III} & \multicolumn{1}{c}{$\alpha$IV} & \multicolumn{1}{c}{$\alpha$V} \\
$E_{L-Glu \ 353}$ & \multicolumn{1}{c}{-24.51} & \multicolumn{1}{c}{-24.50} & \multicolumn{1}{c}{-24.21} & \multicolumn{1}{c}{-23.16} & \multicolumn{1}{c}{-22.92} & \multicolumn{1}{c}{-22.55} & \multicolumn{1}{c}{-27.37} & \multicolumn{1}{c}{-21.00} & \multicolumn{1}{c}{-2.24} & \multicolumn{1}{c}{-24.92} \\  
$E_{L-Arg \ 394}$ & \multicolumn{1}{c}{3.24} & \multicolumn{1}{c}{4.40} & \multicolumn{1}{c}{2.45} & \multicolumn{1}{c}{2.56} & \multicolumn{1}{c}{2.55} & \multicolumn{1}{c}{1.01} & \multicolumn{1}{c}{1.04} & \multicolumn{1}{c}{1.76} & \multicolumn{1}{c}{-9.07} & \multicolumn{1}{c}{-0.06} \\ 
$E_{L-His \ 524}$ & \multicolumn{1}{c}{-16.82} & \multicolumn{1}{c}{-16.42} & \multicolumn{1}{c}{-16.45} & \multicolumn{1}{c}{-15.60} & \multicolumn{1}{c}{-16.61} & \multicolumn{1}{c}{-15.86} & \multicolumn{1}{c}{-13.09} & \multicolumn{1}{c}{-11.96} & \multicolumn{1}{c}{-14.01} & \multicolumn{1}{c}{-13.41} \\
\midrule[1pt]
\\
\\
\midrule[1pt] 
       \multicolumn{11}{c}{In Vacuo FMO2-MP2/6-31G(d)} \\
  \midrule[1pt]
  & \multicolumn{5}{c}{{E2-ER}}
  & \multicolumn{5}{c}{{17$\alpha$-E2-ER}} \\
    \midrule[1pt]
  \ \ \ RS & \multicolumn{1}{c}{$\beta$I} & \multicolumn{1}{c}{$\beta$II} & \multicolumn{1}{c}{$\beta$III} & \multicolumn{1}{c}{$\beta$IV} & \multicolumn{1}{c}{$\beta$V} & \multicolumn{1}{c}{$\alpha$I} & \multicolumn{1}{c}{$\alpha$II} & \multicolumn{1}{c}{$\alpha$III} & \multicolumn{1}{c}{$\alpha$IV} & \multicolumn{1}{c}{$\alpha$V} \\
$E_{L-Glu \ 353}$ & \multicolumn{1}{c}{-26.73} & \multicolumn{1}{c}{-27.16} & \multicolumn{1}{c}{-27.50} & \multicolumn{1}{c}{-25.75} & \multicolumn{1}{c}{-23.49} & \multicolumn{1}{c}{-24.56} & \multicolumn{1}{c}{-28.16} & \multicolumn{1}{c}{-25.48} & \multicolumn{1}{c}{-2.64} & \multicolumn{1}{c}{-26.60} \\  
$E_{L-Arg \ 394}$ & \multicolumn{1}{c}{3.51} & \multicolumn{1}{c}{4.61} & \multicolumn{1}{c}{2.90} & \multicolumn{1}{c}{2.87} & \multicolumn{1}{c}{2.44} & \multicolumn{1}{c}{0.73} & \multicolumn{1}{c}{0.98} & \multicolumn{1}{c}{2.39} & \multicolumn{1}{c}{-9.87} & \multicolumn{1}{c}{-0.28} \\ 
$E_{L-His \ 524}$ & \multicolumn{1}{c}{-17.59} & \multicolumn{1}{c}{-16.84} & \multicolumn{1}{c}{-17.16} & \multicolumn{1}{c}{-15.88} & \multicolumn{1}{c}{-16.86} & \multicolumn{1}{c}{-16.35} & \multicolumn{1}{c}{-13.24} & \multicolumn{1}{c}{-12.21} & \multicolumn{1}{c}{-14.04} & \multicolumn{1}{c}{-13.94} \\  
  \bottomrule[1pt]   
\end{tabular}
\begin{tablenotes}
      \footnotesize
      \item $L$: Ligand. All energies at MP2/6-31G(d) level of theory in kcal/mol.
\end{tablenotes}
\label{Table IV}
  \end{threeparttable}
      \end{center}
\end{table*}

In Figure 5 (bottom), cluster separation for the C$\alpha$ fluctuations of the binding site is displayed. Cluster 1bs is mainly occupied by E2-ER, while cluster 2bs by apo-ER; 17$\alpha$-E2-ER is distributed evenly between both clusters. Obviously, the essential 2D space is smaller because the motions are more restricted than those of entire receptor. By observing the dynamic behavior and the small conformational differences in the models, it is possible to argue that the most active ligand (e.g. 17$\beta$-Estradiol) causes significant changes in the apo-receptor that eventually are translated into biochemical responses. The above argument would explain the fact that 17$\alpha$-E2-ER exhibits, in the binding site, features of both "antipode structures" represented by E2-ER and apo-ER, and is distributed between both clusters Figure 5 (bottom right).

Probability distributions for the backbone C$\alpha$ fluctuations in combined MD trajectory along several principal components are shown in Figure 7. In order to provide insights about the three PCA modes of motion, movies to visualize pseudo-trajectories of the models (download source) were created with the ambertools cpptraj module. The comparison of the top right panel in figure 7 with the third PCA mode movies, taking into account the overlap of the distributions, allows us to ignore the particularities of the motion displayed in the movies and focus on its global features. Consequently, comparing the PCA modes of motion and the respective probability distributions we could conclude the following: (i) A greater plasticity in the movement of apo-ER with respect to both holo-ER. The backbone rigidity in movement is induced by the ligand (second PCA mode movies vs. top middle panel). (ii) A wiggling motion through the principal axis of the estrogen receptor that occurs at apo-ER and 17$\alpha$-E2-ER. This movement has a slightly smaller amplitude in apo-ER (first PCA mode movies vs. top left panel). An analogous analysis at the binding site, including the respective PCA mode movies, showed two types of movement of the ligand; a jiggling motion and a seesaw motion around a principal axis lying in the plane of the molecule. The amplitude of motion was larger in 17$\alpha$-E2 than in E2, which results in a significant disturbance of the binding site. In fact, the interaction with His 524 is continuously hindered by the strong movement of 17$\alpha$-E2. The normal substrate E2, fits perfectly to the binding site and its movement, to some extent, harmonizes with the whole receptor. In general, when comparing the models we can affirm that apo form show a greater plasticity in its movement, and that E2-ER has less conformational freedom.

\begin{figure*} [!htp]
\begin{center}
\includegraphics[width=0.95\linewidth]{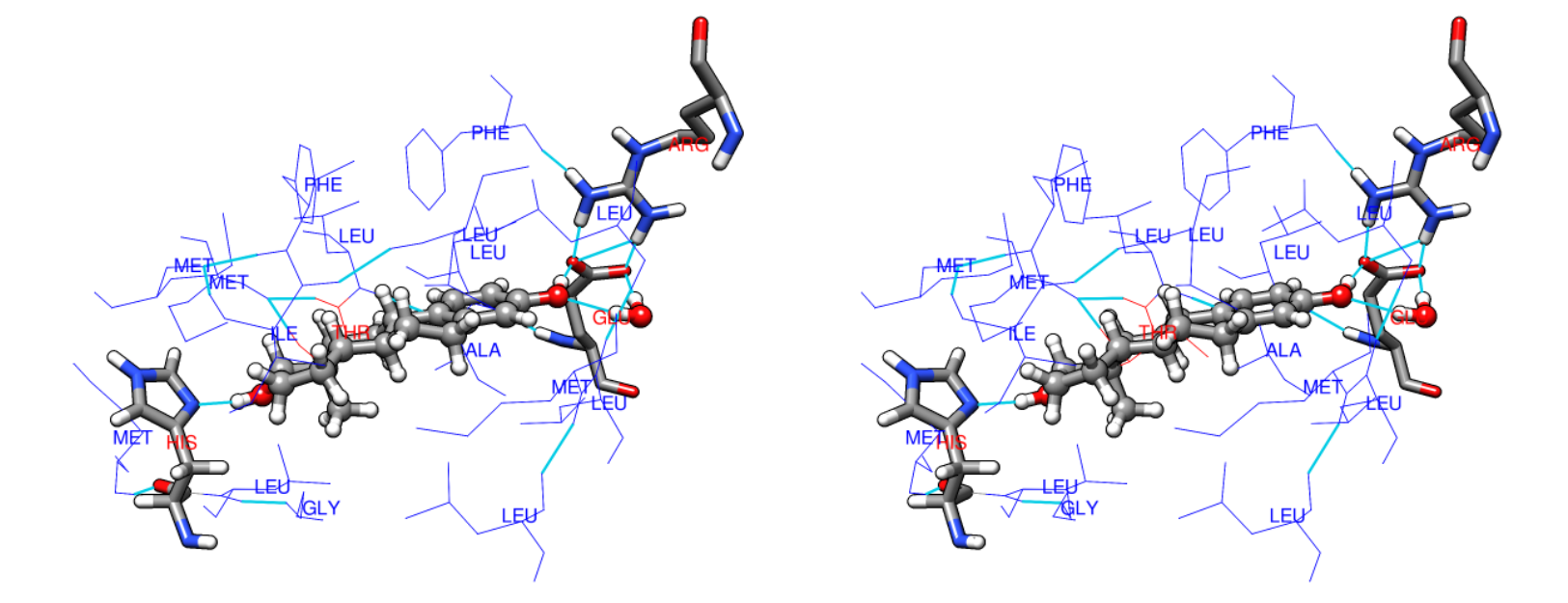}
\end{center}
    \caption{Stereo view of estrogen receptor binding site: 17$\beta$-estradiol (ball and stick) is shown surrounded by a hydrophobic pocket (blue wire), water molecule (ball and stick) and important amino acid residues (stick): His 524, Glu 353 and Arg 394. H-bond (cyan). The image corresponds to $\beta$I representative structure.}
    \label{Figure8}
\end{figure*}

\begin{figure*} [!htp]
\begin{center}
\includegraphics[width=0.95\linewidth]{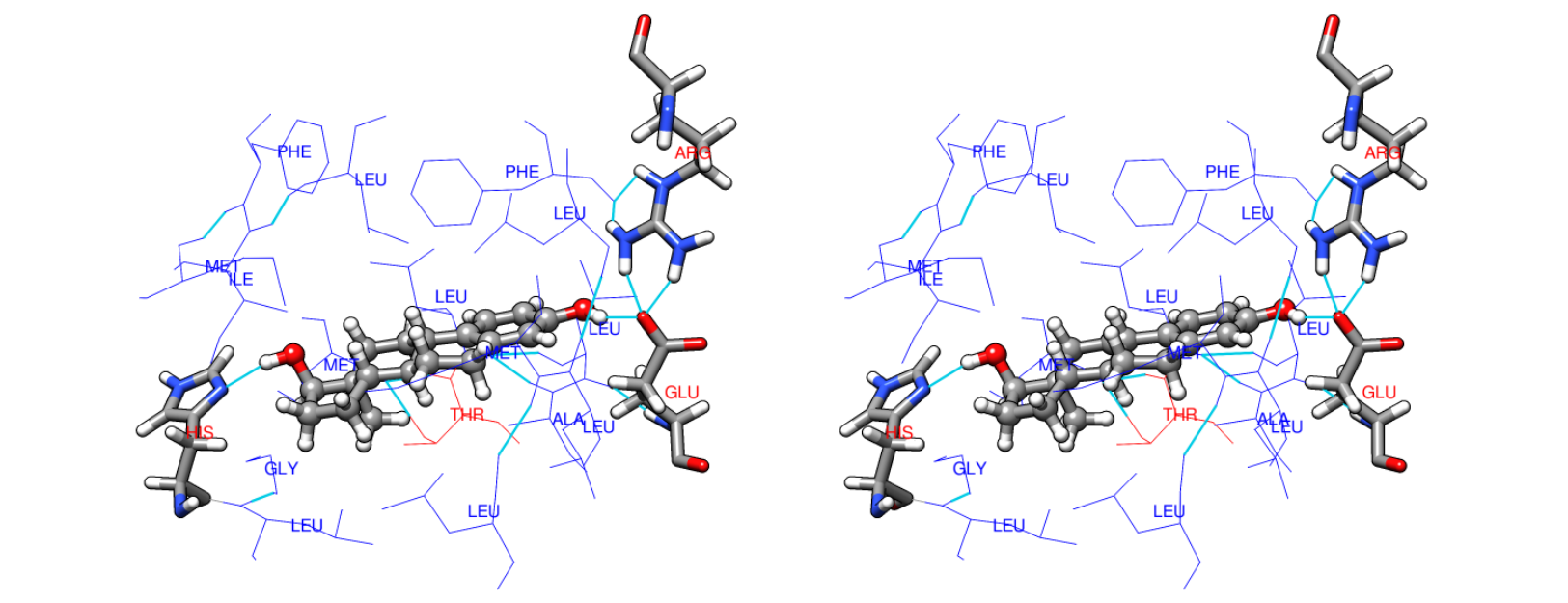}
\end{center}
    \caption{Stereo view of estrogen receptor binding site: 17$\alpha$-estradiol (ball and stick) is shown surrounded by a hydrophobic pocket (blue wire) and important amino acid residues (stick): His 524, Glu 353 and Arg 394. H-bond (cyan). The image corresponds to $\alpha$I representative structure.}
    \label{Figure9}
\end{figure*}

\begin{figure*} [!htp]
\begin{center}
\includegraphics[width=0.95\linewidth]{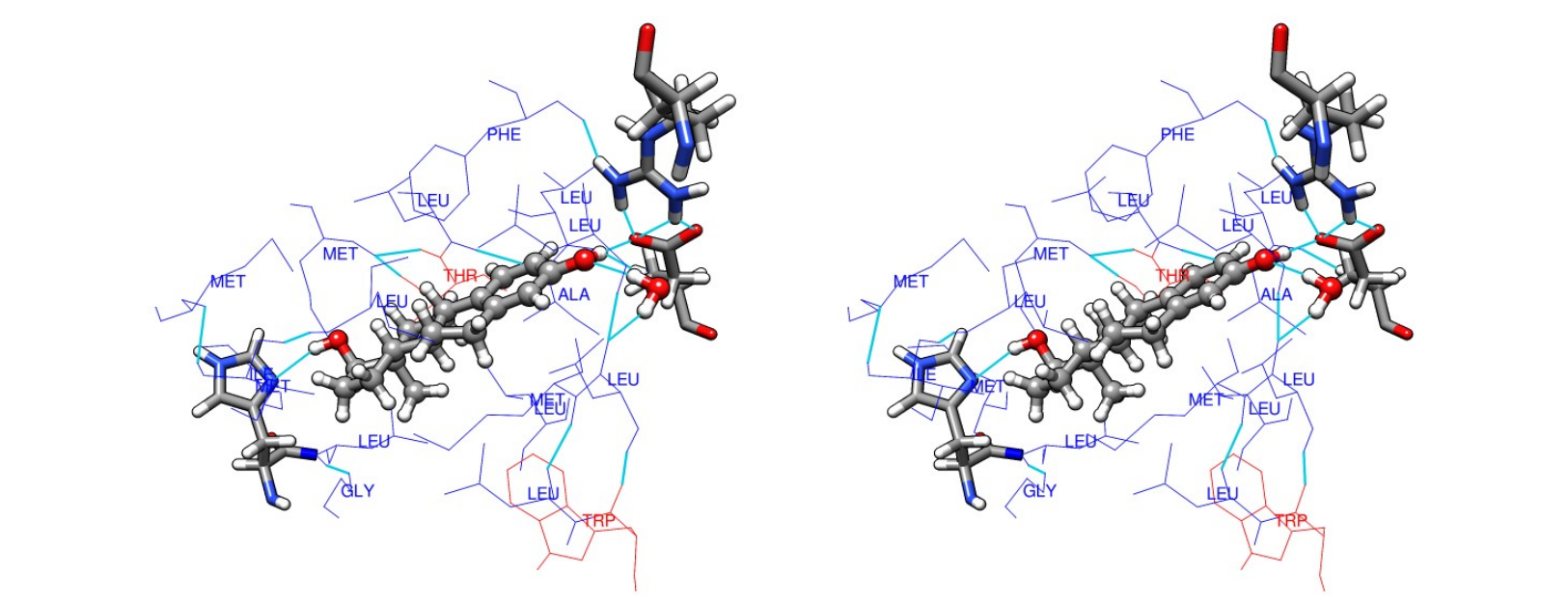}
\end{center}
    \caption{Stereo view of estrogen receptor binding site: 17$\alpha$-estradiol (ball and stick) is shown surrounded by a hydrophobic pocket (blue wire), water molecule (ball and stick) and important amino acid residues (stick): His 524, Glu 353 and Arg 394. H-bond (cyan). The image corresponds to $\alpha$III representative structure.}
    \label{Figure10}
\end{figure*}

\subsection{FMO Interaction Energies}

From MD simulations, 7500 snapshots of each model were grouped into five clusters. Thus, five representative structures (RS) of each cluster were obtained. The interaction energies at MP2/6-31G(d) are shown in Table III. The L-ER interaction energy ($E_{int}$) values are negative which, at least at this level of calculation, would indicate a certain stability of the systems.

In a FMO study of HER$\alpha$ LBD in complex with 17$\beta$-estradiol (PDB code 1ERE), the model included 241 amino acid residues, one water molecule which directly mediates E2-ER binding (where the hydrogen bonded water molecule was incorporated in the receptor) and E2. The binding energy was estimated from: $\Delta E_{Binding} = E_{E2-ER} - (E_{ER} + E_{E2})$. The total energies (E) were considered in the calculation and the geometries of ER and E2 were fixed in those found in E2-ER model. The reported binding energy was -37.65 kcal/mol at FMO2-RHF/STO-3G level of theory \cite{21}. In another FMO study with the same model system, FMO2 interaction energy between E2 and ER was calculated using HF and MP2 methods with several basis sets \cite{60}. The calculated interaction energy was -40.26 kcal/mol at FMO2-RHF/6-31G(d) and -123.73 kcal/mol at FMO2-MP2/6-31G(d) level of theory. The large interaction energy difference between the RHF and the MP2 methods is due to dispersion interaction, which can only be described by electron correlation methods. In general, charged and polarized amino acid residues interact either strongly or weakly with the ligand, while hydrophobic residues contribute to weak interactions. The sum of these weak dispersion interactions makes the difference between both methods. In the present study, interaction energy is reported at the MP2/6-31G(d) level. The results of the calculation of the lower level are excluded.

\begin{table}
 \begin{center}
 \begin{threeparttable}
\caption{Number of water molecules in the binding site}
 \begin{tabular}{cccccc}
 \toprule[1pt] 
 E2-ER-w & $\beta$I & $\beta$II & $\beta$III & $\beta$IV & $\beta$V \\
 water & 1 & 1 & 1 & 1 & 3 \\
 \midrule[1pt]
 17$\alpha$-E2-ER-w & $\alpha$I & $\alpha$II & $\alpha$III & $\alpha$IV & $\alpha$V \\
 water & 0 & 2 & 1 & 2 & 1 \\
 \midrule[1pt]
 ER-w & I & II & III & IV & V \\
 water & 7 & 8 & 6 & 3 & 2 \\
 \bottomrule[1pt]
 \end{tabular}
\label{Table V}
  \end{threeparttable}
  \end{center}
\end{table}

In holo models we observe that Glu 353 and His 524 present a strong electrostatic interaction with the ligand (Table IV), and it is known that together with Arg 394, they form a hydrogen bond network with E2. A water molecule is similarly responsible for yet another stabilizing hydrogen bond between E2 and the apo-ER \cite{22,60}. Many binding site hydrophobic residues are stabilized (attractive interaction) through dispersion interactions with the ligand; the MP2 electron correlation is essential to characterize these interactions, whose function is possibly to accommodate the substrate. Relevant hydrophobic residues at the binding site are: Leu 346, Leu 387, Leu 391, Phe 404 and Leu 525 \cite{61}. Therefore, the interactions between the ligand with these functionally important molecules, together with the hydrogen bonds network play a key role in the L-ER binding (Figures 8-10).

During the simulation water molecules were trapped at the binding site (Table V), which consists of all residues that have at least one atom within 4 {\AA} from any ligand atom in L-ER-w models. This generally gives a good representation of the important residues in the binding pocket of a protein. The number of water molecules varies according to the representative structure and the L-water interactions can be either attractive or repulsive. Focusing on the more populated clusters, which in general terms determine the biological response, the $\alpha$I representative structure lacks a water molecule in the binding site and possibly does not fulfill a basic condition for an adequate attachment of the ligand, and therefore for a suitable biological function. While $\alpha$III contains a water molecule, it represents only 7.2\% of the population of 17$\alpha$-E2-ER. The binding site of $\alpha$V also contains a water molecule, but the phenolic A ring hydroxyl group points in another direction, and probably is not a biologically viable structure. This aspect is interesting since it is possible to infer that not all the structures that present affinity could present efficacy, and therefore, would not be relevant to biological function. That is, only a subset of the conformers population should be biologically functional.

Evidently, the problem arises from the 17$\alpha$-hydroxyl group-His 524 interaction and propagates to the rest of the system. By pointing the $C_{17}$-OH group (D ring) in a direction contrary to that of the normal substrate, it disturbs the anchorage of the molecule in the binding site. This is supported by previous discussion of the PCA modes of motion and justified, to some extent, by the fact that the 17$\alpha$-E2-His 524 interaction energy is on average $\approx$ 1.5 kcal/mol lower than E2-His 524 interaction energy.

Even though both ligands display affinity for the receptor (E2 in greater measure than 17$\alpha$-E2), the population of probably biologically viable conformations is lower in 17$\alpha$-E2 than in E2, and this should be considered when speculating theoretically about the possible action of a ligand on a receptor.

\section{Conclusions}

The molecular interactions between 17$\beta$-estradiol, 17$\alpha$-estradiol and human estrogen receptor $\alpha$ were calculated, and from these the L-ER interaction energy ($E_{int}$) was obtained. The FMO2-MP2/6-31G(d) $E_{int}$ was of -88.52 kcal/mol for E2-ER-w and -78.73 kcal/mol for 17$\alpha$-E2-ER-w models. In vacuo, $E_{int}$ was of -91.69 kcal/mol and -82.00 kcal/mol for E2-ER and 17$\alpha$-E2-ER models, respectively. The results obtained for 17$\alpha$-estradiol clearly suggests that it tends to form a complex with the receptor.

In general, attractive dispersion interactions were observed between ligands and hydrophobic residues; these interactions play an important role in stabilizing E2 and 17$\alpha$-E2 at the binding site. Water molecules were found at the binding site of all representative structures, except $\alpha$I. Strong attractive electrostatic interactions were observed between the ligands and the following charged/polarized residues: Glu 353, His 524. These residues tend to be located at the ends of the ligands, close to the OH groups of the A and D rings.

The C$_{17}$ epimer of E2 disturbs the binding site due to its strong motion, while E2 fits perfectly to the binding site and its motion of lower amplitude, in a sense, synchronizes with the whole receptor. Perhaps this feature of the normal substrate is a necessary condition for biological function. Another important requirement relates to the number of water molecules at the binding site. Therefore, negative values in $E_{int}$ is a necessary but not sufficient condition since, it is also necessary to consider the conformers population that fulfill all the requirements that ensure a biological response.

\section*{Acknowledgements}

I wish to express my gratitude to arXiv repository and to each and every person who has contributed to the development and maintenance of free and open source software.

\bibliographystyle{ieeetr}
\bibliography{manuscrito}

\end{document}